\documentclass[twocolumn,traditabstract]{aa}
\usepackage{txfonts}
\usepackage{color}
\usepackage{graphicx}
\usepackage{natbib}
\usepackage{multirow}
\usepackage[colorlinks=true, allcolors=blue]{hyperref}

\begin{document}

   \title{Dust mass in protoplanetary disks with porous dust opacities}
   \author{Yao Liu\inst{1,2}
          \and
     		  Hélène Roussel\inst{3}
     		  \and
               Hendrik Linz\inst{4}
               \and
     		  Min Fang\inst{2}
     		  \and
     		  Sebastian Wolf\inst{5}
     		  \and
     		  Florian Kirchschlager\inst{6}
     		  \and
               Thomas Henning\inst{4}
	     	  \and
	     	  Haifeng Yang\inst{7}
	     	  \and
	     	  Fujun Du\inst{2}
	     	  \and
			  Mario Flock\inst{4}
			  \and
	     	  Hongchi Wang\inst{2}
          }

    \institute{School of Physical Science and Technology, Southwest Jiaotong University, Chengdu 610031, China; \\
	           \email{yliu@swjtu.edu.cn}
            \and
            Purple Mountain Observatory, Chinese Academy of Sciences, 10 Yuanhua Road, Nanjing 210023, China
     		\and
	    	Institut d’Astrophysique de Paris, Sorbonne Université, CNRS (UMR7095), 75014 Paris, France
             \and
             Max-Planck-Institut f\"ur Astronomie, K\"onigstuhl 17, D-69117 Heidelberg, Germany
	     	\and
             Institut f\"ur Theoretische Physik und Astrophysik, Christian-Albrechts-Universit\"at zu Kiel, Leibnizstr. 15, 24118 Kiel, Germany
	     	\and
	     	Sterrenkundig Observatorium, Ghent University, Krijgslaan 281-S9, B9000 Gent, Belgium
	     	\and
	     	Institute for Astronomy, School of Physics, Zhejiang University, 866 Yu Hang Tang Road, Hangzhou, Zhejiang 310027, China}
        \authorrunning{Liu et al.}
        \titlerunning{Dust mass in protoplanetary disks with porous dust opacities}

   \abstract{Atacama Large Millimeter/submillimeter Array surveys have suggested that protoplanetary disks are not massive enough to form the 
   known exoplanet population, under the assumption that the millimeter continuum emission is optically thin. In this work, we investigate 
   how the mass determination is influenced when the porosity of dust grains is considered in radiative transfer models. The results show 
   that disks with porous dust opacities yield similar dust temperature, but systematically lower millimeter fluxes compared to disks 
   incorporating compact dust grains. Moreover, we recalibrate the relation between dust temperature and stellar luminosity for a wide range of 
   stellar parameters, and calculate the dust masses of a large sample of disks using the traditionally analytic approach. The median dust 
   mass from our calculation is about 6 times higher than the literature result, and this is mostly driven by the different opacities 
   of porous and compact grains. A comparison of the cumulative distribution function between 
   disk dust masses and exoplanet masses show that the median exoplanet mass is about 2 times lower than the median dust mass, if grains are porous,
   and there are no exoplanetary systems with masses higher than the most massive disks. Our analysis suggests that adopting porous dust opacities may alleviate 
   the mass budget problem for planet formation. As an example illustrating the combined effects of optical depth and porous dust opacities on the 
   mass estimation, we conduct new IRAM/NIKA-2 observations toward the IRAS\,04370+2559 disk and perform a detailed radiative transfer modeling of 
   the spectral energy distribution. The best-fit dust mass is roughly 100 times higher than the value from the traditionally analytic calculation. 
   Future spatially resolved observations at various wavelengths are required to better constrain the dust mass.}


   \maketitle

\section{Introduction}

Dust grains, though occupying only about 1\% of the total disk mass, is a crucial ingredient that influences many aspects of disk evolution and 
planet formation \citep[e.g.,][]{Natta2007,Birnstiel2023}. First, dust opacity dominate over gas opacity, therefore dust grains play an important role 
in setting the thermal and geometrical structure of disks. Secondly, dust particles provide the surface area on which complex chemical reactions take 
place \citep[e.g.,][]{Garrod2006,Henning2013,Oberg2023}, and also act as carriers to redistribute volatile species within the disk via dust diffusion, 
settling and radial drift \citep[e.g.,][]{Krijt2016,Stammler2017,Eistrup2022}. Finally, dust grains are the building blocks for the formation of 
planetesimals, terrestrial planets, and the cores of giant planets. Consequently, the total amount of dust content is one of the paramount 
properties that characterize the potential for planet formation.

Estimating dust mass is commonly accomplished by millimeter continuum observations. Thanks to the high sensitivity of the Atacama Large 
Millimeter/submillimeter Array (ALMA), a large sample of about 1,000 disks located in nearby star-forming regions have been observed at millimeter 
wavelengths, \citep[e.g.,][]{Ansdell2016,Barenfeld2016,Pascucci2016,Cazzoletti2019,Tazzari2021,Grant2021}. Assuming the emitting dust is 
optically thin and isothermal, the measured flux density $F_{\nu}$ can be converted into dust masses via the analytic formula  
\begin{equation}
M_{\rm dust} = \frac{F_{\nu}D^2}{\kappa_{\nu}B_{\nu}(T_{\rm dust})},
\label{eqn:mana}
\end{equation}
where $B_{\nu}(T_{\rm dust})$ stands for the Planck function given at the observed frequency $\nu$ and dust temperature $T_{\rm dust}$, $D$ 
refers to the distance to the object, and $\kappa_{\nu}$ is the mass absorption coefficient. Based on Eq.~\ref{eqn:mana}, \citet{Andrews2020} build the cumulative distribution function (CDF) of $M_{\rm dust}$ for 887 disks, and find that less than ${\sim}10\%$ disks have enough material to produce our solar system or its analogs in the exoplanet population. Similarly, \citet{Manara2018} find that exoplanetary systems masses are comparable or even higher than the most massive disks with ages of ${\sim}1\,{-}\,3\,\rm{Myr}$. Although the discrepancy between the disk and exoplanets mass distributions is mitigated by accounting for observational selection and detection biases \citep{Mulders2021}, the findings by \citet{Andrews2020} and \citet{Manara2018} naturally raise a conundrum that protoplanetary disks do not have enough mass to make planetary systems. This puzzle is referred as the ``mass budget problem'' of planet formation in the literature.

Proposed solutions have been considered in three scenarios. The first scenario suggests that planet formation might begin at earlier 
stages of disk evolution than previously thought. The prevalence of substructures in protostellar disks is the supporting 
evidence \citep{SeguraCox2020,Sheehan2020,Ohashi2023,Hsieh2024}, since disk substructures may be created by planets \citep{Kley2012,Paardekooper2023}. 
Moreover, young disks in the Class 0/I phases are generally more massive than Class II disks \citep{Tychoniec2018,Tychoniec2020}, providing more material 
for planet formation. The second scenario assumes that the disk acts as a conveyor belt that transports material from the environment to the central 
star \citep{Manara2018,Gupta2023,Gupta2024}. Thus, the total amount of material available for planet formation actually exceeds the observed value. The third proposal 
argues that $M_{\rm dust}$ estimated using Eq.~\ref{eqn:mana} is merely a lower limit \citep[e.g.,][]{Ballering2019}, because protoplanetary disks 
are not necessarily optically thin at millimeter wavelengths \citep[e.g.,][]{Liuy2017,Rilinger2023,Xin2023}. Recently, \citet{Savvidou2024} show 
that the early formation of giant planets is expected to create pressure bumps exterior to the planetary orbit, which will trap the inward drifting 
dust. The trapped dust will be largely unaccounted for by the approximation of optically thin emission. Assuming dust grains are compact 
spheres, \citet{Liuy2022} demonstrate that the disk outer radius, inclination, and true dust mass are most important to create optically thick 
regions, resulting in mass underestimations from a few to hundreds times. 

Theoretical studies show that dust grains in protoplanetary disks might be porous in the process of coagulation and growth \citep[e.g.,][]{Dominik1997,Wurm1998}, and grain growth via porous aggregates can overcome the radial drift barrier to form planetesimals \citep{Okuzumi2012,Kataoka2013,Michoulier2024}. The existence of porous dust grains is observationally supported. Near-infrared scattered light observations show that the polarization phase functions are consistent with model predictions incorporating micron-sized porous dust grains \citep[e.g.,][]{Stolker2016,Ginski2023,Tazaki2023}. A detailed analysis of multiwavelength continuum and millimeter polarization observations of the HL\,Tau disk indicates that dust grains are porous, and the porosity ranges from $70\%$ to $97\%$ depending on the best-fit grain size \citep{Zhang2023}.  

The absorption/scattering coefficients of porous dust grains are different from those of compact spherical particles \citep[e.g.,][]{Kirchschlager2014,Ysard2018,Kirchschlager2019}, which have a direct impact on the radiative transfer process in protoplanetary disks. In this work, using self-consistent radiative transfer models, we investigate how porous dust opacities influence the dust temperature, emergent millimeter flux and therefore the estimation of $M_{\rm dust}$. The setup of the radiative transfer models and the resulting trends are presented in Sect.~\ref{sec:rtmodel}. In Sect.~\ref{sec:exoplanet}, we apply porous dust opacities to the calculation of $M_{\rm dust}$ for a large number of disks, and compare the CDF between the disk sample and the exoplanet population investigated by \citet{Mulders2021}. As an application, we model the spectral energy distribution (SED) of the IRAS\,04370+2559 disk in Sect.~\ref{sec:iras04370}. The paper concludes with a summary in Sect.~\ref{sec:summary}.

\section{Radiative transfer models}
\label{sec:rtmodel}

In this section, we first give an introduction about the setup of the radiative transfer model, and then build a grid of models by sampling a few 
key parameters that have prominent effects on the dust temperature and millimeter flux. 

\subsection{Dust density structure}
We consider a flared disk that includes two distinct dust grain populations, i.e., a small grain population (SGP)
and a large grain population (LGP). We fix the disk inner radius $R_{\rm in}$ to 0.1\,AU that is close to the dust 
sublimation radius of a typical T Tauri star. The SGP occupies a small fraction of the dust mass, $(1-f)\,M_{\rm dust}$, and 
its scale height follows a power law
\begin{equation}
h = h_{100}\times\left(\frac{R}{100\,\rm{AU}}\right)^\Psi.
\label{eq:heightgas}
\end{equation}
The flaring index is denoted with $\Psi$, and $h_{100}$ refers to the scale height at a radial distance of  
$R\,{=}\,100\,\rm{AU}$. On the contrary, the LGP which has a mass of $f\,M_{\rm dust}$ dominates the dust mass, and is concentrated 
close to the midplane with a scale height of $\Lambda\,h$. The degree of dust settling is characterized by setting 
$\Lambda\,{=}\,0.2$ \citep{Andrews2011}. To distribute the mass for the SGP and LGP, we adopt $f\,{=}\,0.85$ that is a typical value found 
from multiwavelength modeling of protoplanetary disks \citep[e.g.,][]{Andrews2011,vanderMarel2018,Schwarz2021,Zhang2021}. The dust surface density is 
assumed to be a power law with an exponential taper
\begin{equation}
\Sigma(R)\,{=}\,\Sigma_{c}\left(\frac{R}{R_{c}}\right)^{-\gamma}{\rm exp}\left[-\left(\frac{R}{R_{c}}\right)^{2-\gamma}\right],
\label{eqn:sigma}
\end{equation}
where $\gamma$ is the gradient parameter, and $R_{\rm c}$ is a characteristic radius. This is the similarity solution for 
disk evolution, which assume that the viscosity has a power-law radial dependence and is independent of time \citep{Lynden-Bell1974}. 
The proportionality factor $\Sigma_{\rm c}$ is determined by normalizing the total dust mass $M_{\rm dust}$.
We truncate the disk at an outer radius of $8\,R_{c}$. The dust volume density is parameterized as 
\begin{equation}
\rho_{\rm{SGP}}(R,z)\,{=}\,\frac{(1{-}f)\Sigma(R)}{\sqrt{2\pi}h}\,\exp\left[-\frac{1}{2}\left(\frac{z}{h}\right)^2\right], \\
\label{eqn:sgp}
\end{equation}
\begin{equation}
\rho_{\rm{LGP}}(R,z)\,{=}\,\frac{f\Sigma(R)}{\sqrt{2\pi}\Lambda h}\,\exp\left[-\frac{1}{2}\left(\frac{z}{\Lambda h}\right)^2\right]. \\
\label{eqn:lgp}
\end{equation}
Table~\ref{tab:parasample} summarizes the parameters of the model.

\begin{table}
\caption{Fixed and varied parameters of the model grid.}
\centering
\linespread{1.4}\selectfont
\begin{tabular}{lcccc}
\hline
Parameter                     &  Min       &    Max         &   Number       &  Sampling  \\
\hline
$T_{\star}\, [\rm K]$         &  3778      & 3778    &     1    &      fixed  \\
$L_{\star}\, [L_{\odot}]$     &  0.86      & 0.86    &     1    &      fixed  \\
$R_{\rm in}\,[\rm{AU}]$       &  0.1       & 0.1     &     1    &      fixed  \\
$\gamma$                      &  1.0       & 1.0     &     1    &      fixed  \\
$\Lambda$                     &  0.2       & 0.2     &     1    &      fixed  \\
$f$                           &  0.85      & 0.85    &     1    &      fixed  \\
$i\,[\circ]$                  &  30        & 30      &     1    &      fixed  \\
$a_{\rm min}^{\rm eff}\, [\mu{\rm m}]$     &  0.01   &  0.01    &   1  &  fixed   \\
$a_{\rm max.SGP}^{\rm eff}\, [\mu{\rm m}]$ &  1.0    &  1.0     &   1  &  fixed   \\   
$a_{\rm max.LGP}^{\rm eff}\, [{\rm mm}]$   &  1.0    &  1.0     &   1  &  fixed   \\
$R_{\rm c}\,[\rm{AU}]$        &  5         & 200     &     10   &      logarithmic \\
$\Psi$                        &  1.05      & 1.25    &     9    &      linear    \\
$h_{\rm 100}\,[\rm{AU}]$      &  4         & 18      &     8    &      linear    \\ 
$M_{\rm dust}\, [M_{\odot}]$  & $1\,{\times}\,10^{-5}$ &  $1\,{\times}\,10^{-2}$  &  13  &  logarithmic \\
\hline
\end{tabular}
\label{tab:parasample}
\end{table}

\begin{figure} 
   \centering
   \includegraphics[width=0.4\textwidth]{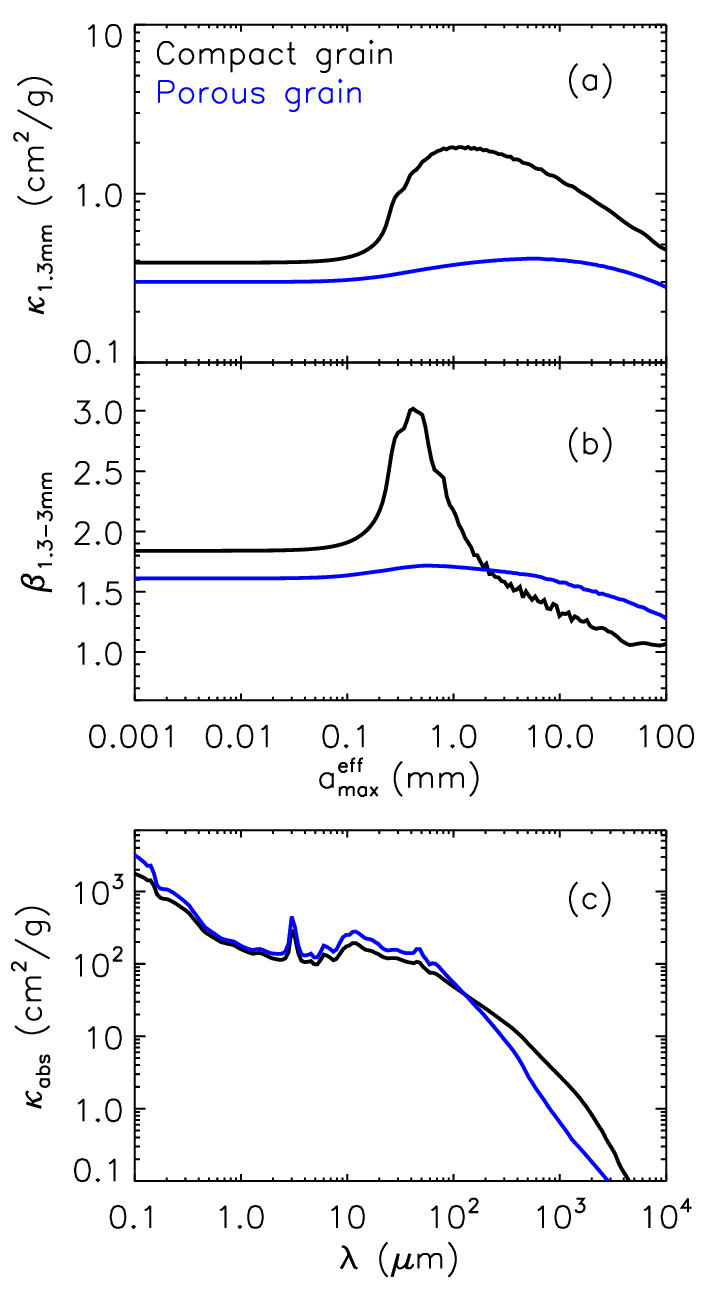}
   \caption{Dust properties considered in this work. {\it Panel (a):} mass absorption coefficient at $\lambda\,{=}\,1.3\,\rm{mm}$ 
   as a function of $a_{\rm max}^{\rm eff}$. {\it Panel (b):} opacity slope $\beta_{1.3-3\rm{mm}}$ between 1.3\,mm and 3\,mm. 
   The grain size distribution follows a power law ${\rm d}n(a_{\rm eff})\,{\propto}\,{a_{\rm eff}^{-3.5}} {\rm d}a_{\rm eff}$
   with a fixed $a_{\rm min}^{\rm eff}\,{=}\,0.01\,\mu{\rm m}$. {\it Panel (c):} mass absorption coefficient as a function of wavelength 
   for $a_{\rm max}^{\rm eff}\,{=}\,1\,\rm{mm}$. The black lines show the result for compact grains, whereas the case for the porous grains 
   with a porosity of $\mathcal{P}\,{=}\,0.8$ is indicated with blue lines.} 
   \label{fig:dustopac}
\end{figure} 

\subsection{Dust properties}
\label{sec:dustmodel}

The dust ensemble is composed of water ice \citep{Warren2008}, astronomical silicates \citep{Draine2003}, troilite \citep{Henning1996}, and refractory organic material \citep{Henning1996}, with volume fractions being 36\%, 17\%, 3\%, and 44\%, respectively. The optical constants of each individual are mixed using the Bruggeman mixing rule \citep{Bruggeman1935}. The resulting optical constants are referred as the DSHARP dust model \citep{Birnstiel2018}. The bulk density of the mix is $\rho_{\rm s.compact}\,{=}\,1.675\,\rm{g/cm^{3}}$. To have porous grains, we adopt the effective medium theory (EMT), and mix the DSHARP dust composition with vacuum using the Bruggeman mixing rule. The mixed complex refractive indices are used to calculate dust absorption/scattering properties with the Mie theory and \texttt{OpTool} \citep{Dominik2021}. The porosity $\mathcal{P}$ controls the volume fraction of vacuum. We choose $\mathcal{P}\,{=}\,0.8$, a value consistent with current observations of protoplanetary disks \citep[e.g.,][]{Zhang2023} and also derived for the cometary nucleus 67P/Churyumov–Gerasimenko \citep{Kofman2015,Jorda2016}. The bulk density of the porous dust grains is $\rho_{\rm s.porous}\,{=}\,(1\,{-}\,\mathcal{P})\,\rho_{\rm s.compact}\,{=}\,0.335\,\rm{g/cm^{3}}$. Note that compact dust particles can be considered as a limiting case of $\mathcal{P}\,{=}\,0$.

We define an effective radius that is the radius of a volume-equivalent solid sphere, $a_{\rm eff}\,{=}\,a\,(1-\mathcal{P})^{1/3}$, where $a$ is the grain size. In this work, we compare compact dust opacities with porous dust opacities assuming that both types of dust grains have the same effective radius $a_{\rm eff}$. In the literature, the characteristic radius $a_{\rm c}\,{=}\,a\,(1-\mathcal{P})$ is another choice for describing the size-dependent dust opacities \citep{Tazaki2019,Zhang2023}. For $\mathcal{P}\,{=}\,0.8$, $a_{\rm eff}$ and $a_{\rm c}$ differ by a factor of ${\sim}\,3$. The effective grain size distribution follows the power law ${\rm d}n(a_{\rm eff})\,{\propto}\,{a_{\rm eff}^{-3.5}} {\rm d}a_{\rm eff}$ with a minimum effective grain size fixed to $a_{\rm{min}}^{\rm eff}\,{=}\,0.01\,\mu{\rm m}$. For the SGP, the maximum effective grain size is set to $a_{\rm{max}}^{\rm eff}\,{=}\,1\,\mu\rm{m}$. For the LGP, we adopt $a_{\rm{max}}^{\rm eff}\,{=}\,1\,\rm{mm}$ accounting for grain growth commonly identified in protoplanetary disks. The porosity may not be uniform due to the dust coagulation of the SGP into the LGP. However, we focus only on the simplest scenario where the SGP and LGP are assumed to have the same porosity of $\mathcal{P}\,{=}\,0.8$.

The panel (a) of Figure~\ref{fig:dustopac} shows the mass absorption coefficient at 1.3\,mm ($\kappa_{\rm 1.3mm}$) as a function of $a_{\rm{max}}^{\rm eff}$. As can be seen, $\kappa_{\rm 1.3mm}$ of porous grains is lower than that of compact grains \footnote{When the particle sizes are in the Rayleigh limit, fluffy grains can have increased dust opacities at millimeter wavelengths compared to compact grains with the same mass, and the increase is highly dependent on the dust composition \citep[e.g.,][]{Stognienko1995,Henning1996}.}. For $a_{\rm{max}}^{\rm eff}\,{=}\,1\,\rm{mm}$, porous grains feature $\kappa_{\rm 1.3mm}\,{=}\,0.37\,\rm{cm^2/g}$ that is five times lower than compact grains with $\kappa_{\rm 1.3mm}\,{=}\,1.86\,\rm{cm^2/g}$, which will have a direct impact on the mass estimation, see Eq.~\ref{eqn:mana}. The differences in $\kappa_{\nu}$ between porous grains and compact grains are less pronounced in the optical regime, see panel (c) of Figure~\ref{fig:dustopac}. This implies that $T_{\rm dust}$ obtained from the radiative transfer simulation will not differ too much between the two types of grains because most of the stellar energy emits at optical wavelengths. The opacity slope, e.g., $\beta_{\rm 1.3-3mm}$ measured between 1.3\,mm and 3\,mm, behaves differently between compact grains and porous grains. For compact grains, there is a peak around $a_{\rm{max}}^{\rm eff}\,{\sim}\,\lambda/2\pi$ due to the unique feature of Mie interference. For larger grain sizes, the opacity slope monotonically decreases with $a_{\rm{max}}^{\rm eff}$. If the dust emission is optically thin and in the Rayleigh–Jeans tail, the opacity slope is linked to millimeter spectral indices ($\alpha$) via $\beta\,{=}\,\alpha\,{-}\,2$. This is the reason why dust grain sizes can be probed by multiwavelength millimeter observations \citep[e.g.,][]{Ricci2010b,Williams2011}. However, if the dust particles are porous, the grain size becomes difficult to be constrained because the opacity slope is less sensitive to $a_{\rm{max}}^{\rm eff}$, see panel (b) of Figure~\ref{fig:dustopac} or Figure 3 in \citet{Miotello2023}.

As described above, we modeled the porosity as a certain percentage of vacuum inclusion in dust grains by using the EMT, and calculated the dust properties with the Mie theory. A more sophisticated approach is to define the porosity by a size of void inclusion and a volume filling factor \citep[e.g.,][]{Kirchschlager2014,Kirchschlager2019}. In this way, dust properties of porous grains can be calculated with the \texttt{DDSCAT} code \citep{Draine1994,Draine2010} by applying the discrete dipole approximation \citep{Purcell1973}. \citet{Kirchschlager2013} showed that absorption cross sections of micron-sized grains calculated with the Mie theory and \texttt{DDSCAT} code differ for $\lambda\,{<}\,0.2\,\mu{\rm m}$ and $\lambda\,{>}\,20\,\mu{\rm m}$. Comparisons for millimeter-sized grains are difficult, because running the \texttt{DDSCAT} code for grains with large size parameters ($2\pi a/\lambda$) is time-consuming, and the reliability of results needs to be validated as well. Hence, we leave it as a future direction to investigate the difference in the dust temperature between the two approaches, and how it would impact the estimation of dust mass.
  
\subsection{Heating mechanisms}
\label{sec:heating}
Stellar irradiation and viscous accretion are the major heating sources of protoplanetary disks. Viscous heating mainly affects the temperature
distribution in the interior portion of the inner disk, e.g., $R\,{\lesssim}\,2\,\rm{AU}$ \citep{Hartmann2009,Harsono2015}. However, the main disk mass 
reservoir is in the cold outer disk. Therefore, we only consider stellar irradiation in the simulation.

\begin{figure}[!t]
\centering
  \includegraphics[width=0.4\textwidth]{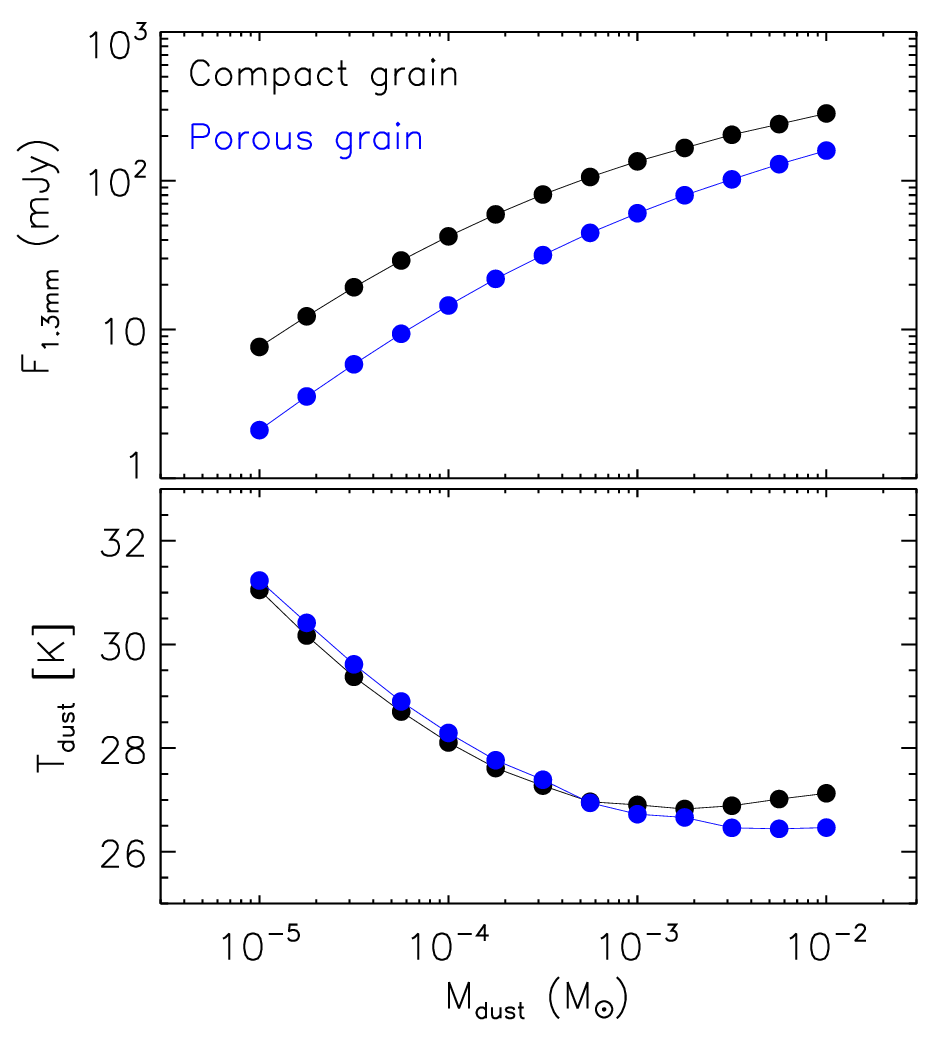}
  \caption{Comparison of $F_{\rm 1.3mm}$ (upper panel) and $T_{\rm dust}$ (bottom panel) between models that use compact dust opacities (black) and porous dust opacities (blue), respectively. In the models, all of other parameters are fixed: $R_{\rm c}\,{=}\,30\,{\rm AU}$, $\Psi\,{=}\,1.15$, $H_{100}\,{=}\,10\,\rm{AU}$.}
\label{fig:compor}
\end{figure}

In Sect.~\ref{sec:iras04370}, we will perform a detailed fitting to the SED of IRAS 04370+2559 that is a T Tauri star located in the Taurus 
star formation region. For convenience, we adopt its stellar luminosity and effective temperature ($L_{\star}\,{=}\,0.86\,L_{\odot}$, 
$T_{\rm{eff}}\,{=}\,3778\,{\rm K}$, see Sect.~\ref{sec:obssed}) in the establishment of the model grid. We take the distance 
of $D\,{=}\,140\,\rm{pc}$ to scale the simulated observable. We note that using different stellar properties will not have a significant 
impact on the trends of $T_{\rm dust}$ and millimeter $F_{\nu}$ with parameters explored in this work. The stellar spectrum is taken 
from the \texttt{BT-Settl} database \citep{Allard2011}, assuming a surface gravity of ${\rm log}\,g\,{=}\,3.5$ and solar metallicity. 

\subsection{Establishment of the model grid}
\label{sec:grid}

To build a grid of models, we explore a few key parameters that are expected to have the most significant impact on $T_{\rm dust}$ and the millimeter flux $F_{\nu}$. 
The flaring index $\Psi$ and scale height $h_{100}$ work together to determine the disk geometry, therefore to alter the total amount of stellar energy absorbed by 
the disk, which in turn affects $T_{\rm dust}$ and $F_{\nu}$. As shown by \citet{Liuy2022}, the true $M_{\rm dust}$, disk size and disk inclination ($i$) are most 
important to create optically thick regions. Accordingly, $M_{\rm dust}$ and $R_{c}$ have an impact on the mass estimation. The disk inclination affects the optical depth along the line of sight, and therefore alters the millimeter $F_{\nu}$. We refer to \citet{Liuy2022} for details about the role of this parameter in determining the dust mass. For simplicity, we fixed $i\,{=}\,30^{\circ}$ throughout this work. 

We sample the above-mentioned four parameters ($\Psi$, $h_{100}$, $M_{\rm dust}$ and $R_{c}$) within reasonable ranges that are consistent with results derived from 
multiwavelength observations and modeling of protoplanetary disks \citep[e.g.,][]{Andrews2011,Kirchschlager2016,Andrews2013,Liuy2019}. For parameters having a large 
dynamical range, the sampling is performed in a logarithmic manner, whereas parameters with a narrow dynamical range are sampled in the linear space. The grid points 
in each dimension are tabulated in Table~\ref{tab:parasample}. There are 9,360 models in total. We separately run the simulations using the compact dust opacities 
and porous dust opacities. 

The well-tested \texttt{RADMC-3D}\footnote{\url{http://www.ita.uni-heidelberg.de/~dullemond/software/radmc-3d/.}} package is invoked to solve the problem of continuum 
radiative transfer \citep{radmc3d2012}. Dust scattering is taken into account since literature works have demonstrated that it can reduce the emission 
from optically thick regions \citep{Zhu2019,Sierra2020}, and influence the millimeter spectral index \citep{Liuh2019}. We first run the thermal 
Monte-Carlo simulation, from which the mass-averaged dust temperature $T_{\rm dust}$ is obtained. Then, we simulate the flux densities from optical to 
millimeter wavelengths and the 0.88\,mm images. The 0.88\,mm images are used to derive an effective disk radius that is the radius at which a given fraction of the 
cumulative flux is contained. Following \citet{Tripathi2017}, we compute the radius encircling 68\% of the total flux, and compare our results with those 
measured on observed millimeter images of Taurus disks \citep{Hendler2020}.

\begin{figure}[!t]
\centering
  \includegraphics[width=0.4\textwidth]{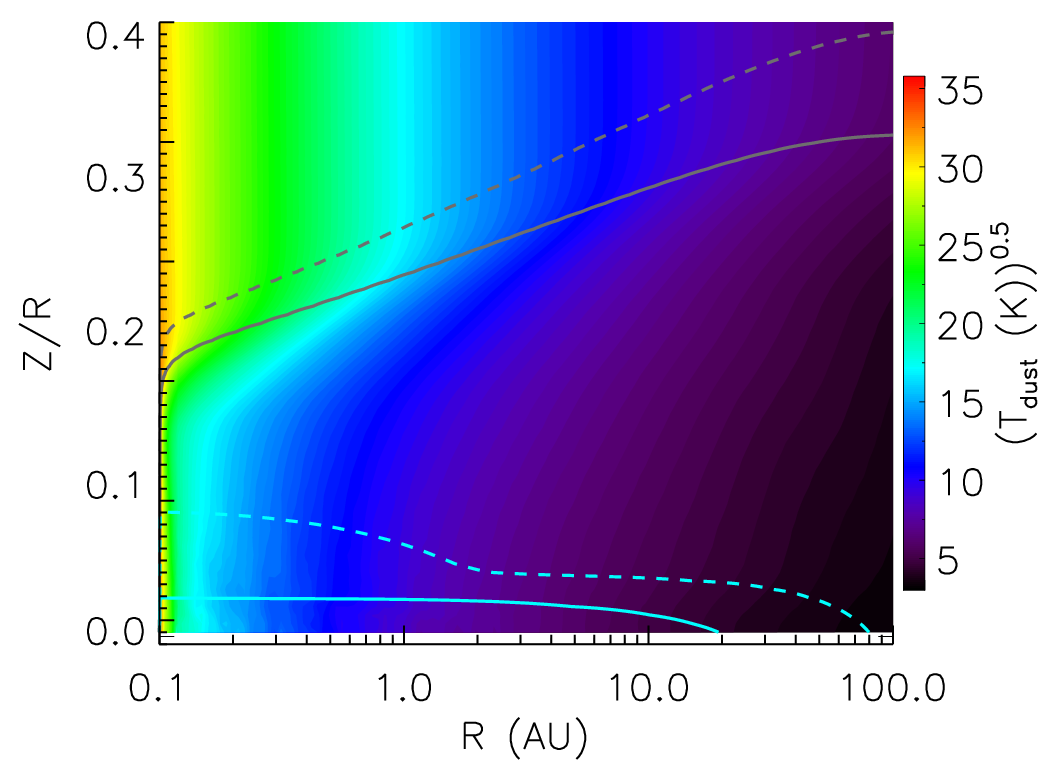}
  \caption{Dust temperature of a representative model with parameters of $M_{\rm dust}\,{=}\,10^{-4}\,M_{\odot}$, $R_{\rm c}\,{=}\,30\,{\rm AU}$, $\Psi\,{=}\,1.15$ and $H_{100}\,{=}\,10\,\rm{AU}$. For a better representation, we show the quantity $\sqrt{T_{\rm dust}}$. The grey solid line refers to the surface of radial optical depth at $\lambda\,{=}\,0.55\,\mu{\rm m}$ being unity, i.e., $\tau_{\rm radial.0.55\,\mu{\rm m}}\,{=}\,1$. This contour marks the disk photosphere where stellar photons are absorbed, and hence plays a crucial role in setting up the temperature structure. The cyan solid line shows the surface of vertical optical depth at $\lambda\,{=}\,1.3\,{\rm mm}$ being unity, i.e., $\tau_{\rm vertical.1.3\,{\rm mm}}\,{=}\,1$. This curve generally reflects the layer above which millimeter continuum emission can reach the observer viewed at a face-on orientation. For comparison, the grey dashed line and cyan dashed line show the same contours for a more massive disk with $M_{\rm dust}\,{=}\,3\times10^{-3}\,M_{\odot}$.}
\label{fig:disktau}
\end{figure}  

\begin{figure}[!t]
\centering
  \includegraphics[width=0.45\textwidth]{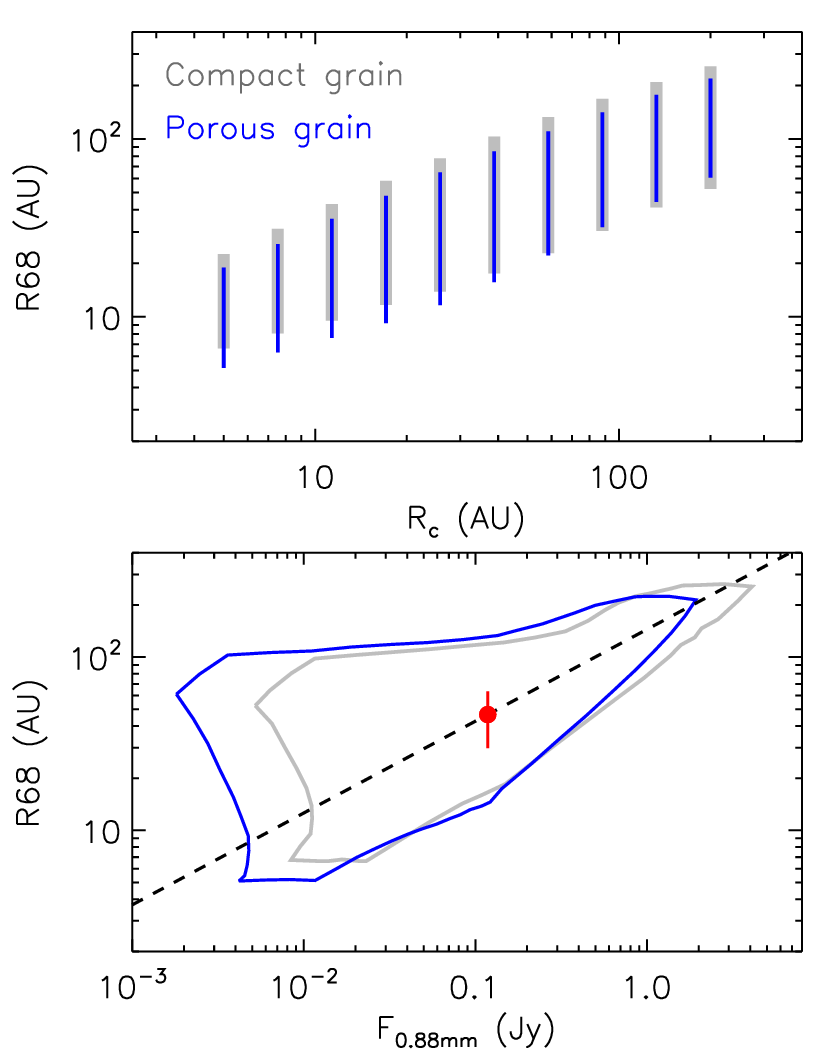}
  \caption{Statistics of the effective disk radius $R68$ from the model grid. {\it Upper panel:} Ranges of $R68$ for each of the sampled $R_{c}$. The blue lines refer to the 
          result when porous dust grains are considered, whereas the grey lines represent the case by using compact dust opacities. The thickness of lines is only 
		  for a better illustration. {\it Bottom panel:} Relation between $R68$ and $F_{\rm 0.88mm}$. The blue curve and grey curve enclose the 
		  regions occupied by the 9,360 models with porous grains and compact grains, respectively. The dashed line shows the relation ${\rm log}\,R68\,{=}\,2.16\,{+}\,0.53\,{\rm log}\,F_{\rm 0.88mm}$ \citep{Hendler2020}. The red dot indicates the expected position of IRAS\,04370+2559 in the diagram, see Sect.~\ref{sec:sedana}.}
\label{fig:r68fnu}
\end{figure}  

\subsection{Results}

Figure~\ref{fig:compor} shows the mass-averaged dust temperature ($T_{\rm dust}$) and flux density at 1.3\,mm ($F_{\rm 1.3mm}$) as a function of $M_{\rm dust}$. 
The difference in $T_{\rm dust}$ between compact dust opacities and porous dust opacities is less than ${\sim}\,1\,{\rm K}$. In self-consistent radiative transfer modeling, only upper disk layers are directly illuminated by the central star, resulting in hot/warm upper disk layers. Generally, such hot/warm disk regions lie above the disk photosphere where the radial optical depth at $0.55\,\mu{\rm m}$ equals to unity, i.e., $\tau_{\rm radial.0.55\,\mu{\rm m}}\,{=}\,1$, which is indicated with the grey lines in Figure~\ref{fig:disktau}. The interior disk regions are indirectly heated by both the radiation that is scattered from above towards the midplane and the reemission of the hot/warm layers. The former heating mechanism is mainly relevant to the optical cross section, while the infrared cross section is important for the latter heating process. As shown in panel (c) of Figure~\ref{fig:dustopac}, porous dust grains feature higher optical cross sections, leading to a lower $T_{\rm dust}$. However, porous dust grains also have larger infrared cross sections. Hot/warm layers will re-emit more infrared emission, and therefore compensate the reduced $T_{\rm dust}$. These facts explain the small difference in $T_{\rm dust}$ between both types of dust grains. Models with porous dust opacities produce systematically lower $F_{\rm 1.3mm}$ than those with compact dust opacities. This is driven by the difference in the dust opacity, see panel (a) of Figure~\ref{fig:dustopac}. For each of the 9,360 models in the grid, we calculate the ratio of $F_{\rm 1.3mm}$ with porous grains to that with compact grains, and derive a median value of 0.4. We observe a less than linear scaling relation between $F_{\rm 1.3mm}$ and $M_{\rm dust}$. Such an outcome can be explained by two reasons. First, the higher optical depth with larger $M_{\rm dust}$ results in a stronger shielding of the inner disk regions, and therefore a lower mass-averaged dust temperature, see the comparison of $\tau_{\rm radial.0.55\,\mu{\rm m}}\,{=}\,1$ surface between two models with different $M_{\rm dust}$ in Figure~\ref{fig:disktau}. Second, with the increasing $M_{\rm dust}$, the millimeter optical depth increases as well, which will hide more disk interior regions and thus their reemitted flux, see the vertical optical depth at a wavelength of 1.3\,mm in Figure~\ref{fig:disktau}. When the disk becomes optically thick, the millimeter emission gradually saturates. Therefore, the difference in $F_{\rm 1.3mm}$ decreases towards a higher $M_{\rm dust}$. 

The disk inner radius is fixed in the model grid. In this case, the disk outer radius defines the radial range in which dust grains are confined, and it significantly affects the optical depth and therefore the mass estimation. From the observational aspect, the disk outer radius is commonly characterized by the effective disk radius from spatially 
resolved images. Previous studies found a strong correlation between the effective disk radius and millimeter flux 
density \citep[e.g.,][]{Tripathi2017,Andrews2018b,Hendler2020}. Using the 0.88\,mm images, we compute the effective disk radius $R68$, i.e., the radius enclosing 
68\% of the total flux, to connect the statistics of our models to those obtained from observations in the literature. The results are presented in 
Figure~\ref{fig:r68fnu}. The effective disk radius depends mainly on the radial optical depth that is determined by the dust opacity and surface 
density (see Eq.~\ref{eqn:sigma}). As can be seen in the upper panel, $R68$ increases with the characteristic radius $R_{\rm c}$. Moreover, the range of 
$R68$ for each sampled $R_{\rm c}$, indicated by the vertical length of the bars, is similar for models with the two types of grains. In the bottom panel 
of the figure, the blue curve and grey curve depict the boundaries within which the 7,920 models with porous dust opacities and compact dust opacities 
occupy in the $R68\,{-}\,F_{0.88{\rm mm}}$ diagram, respectively. Broadly, the blue curve is a shift of the grey curve towards the lower flux direction. 
\citet{Hendler2020} found that $R68$ correlates with $F_{\rm 0.88mm}$ via ${\rm log}\,R68\,{=}\,2.16\,{+}\,0.53\,{\rm log}\,F_{0.88{\rm mm}}$ for Taurus disks. 
Such a correlation, shown with the dashed line, is consistent with the trend revealed by our models. 

\section{Solid mass budget for planet formation}
\label{sec:exoplanet}

Based on self-consistent radiative transfer modeling, we have demonstrated that models with porous dust grains are systematically fainter in the millimeter, but 
the dust temperature is similar to that with compact grains. These facts have a consequence on the solid mass budget for planet formation. In this section, we calculate 
$M_{\rm dust}$ for a large sample of disks using Eq.~\ref{eqn:mana}, and compare the CDF of $M_{\rm dust}$ with that of exoplanet masses. 

\citet{Manara2023} collected $0.88\,\rm{mm}$ and/or $1.3\,\rm{mm}$ flux densities for a large number of nearby disks. 
Assuming $\kappa_{\nu}\,{=}\,2.3(\nu/230\,\rm{GHz)\,cm^2/g}$ and a constant $T_{\rm dust}\,{=}\,20\,\rm{K}$, they investigated the statistics of $M_{\rm dust}$ 
calculated with Eq.~\ref{eqn:mana}. In this work, we focus on the 718 disks with a similar age (i.e., 1\,$-$\,3\,Myr) in their sample, which are located in the 
Taurus, Lupus, Chamaeleon and Ophiuchus molecular clouds. The stellar luminosity of these sources spans a large range of 
$10^{-3}\,L_{\odot}\,{\lesssim}\,L_{\star}\,{\lesssim}\,100\,L_{\odot}$, therefore taking a constant $T_{\rm dust}$ is not appropriate. 
\citet{Andrews2013} showed that $T_{\rm dust}$ mainly depends on $L_{\star}$, and derived a relation $T_{\rm dust}\,{=}\,25\,(L_{\star}/L_{\odot})^{0.25}\,\rm{K}$. 
Nevertheless, their results are based on radiative transfer models with a limited range of $0.1\,L_{\odot}\,{\lesssim}\,L_{\star}\,{\lesssim}\,100\,L_{\odot}$. 
The scaling of $T_{\rm dust}$ with $L_{\star}$ in the lower stellar mass regime was further investigated by \citet{vanderPlas2016} and \citet{Hendler2017}, and  
the relation is found to be flatter $T_{\rm dust}\,{=}\,22\,(L_{\star}/L_{\odot})^{0.16}\,{\rm K}$ \citep{vanderPlas2016}. 

In order to homogenize our analysis, we re-calibrate the $T_{\rm dust}\,{-}\,L_{\star}$ relation from massive young stars all the way down to the brown 
dwarf regime by creating a grid of radiative transfer models. The modeling framework is identical to that described in Section~\ref{sec:rtmodel},
and porous dust opacities are used in the simulation. We first sample 34 values of $L_{\star}$ that are logarithmically distributed in the range 
of [$10^{-3}\,L_{\odot}$, $100\,L_{\odot}$]. Then, we interpolate the 2.5\,Myr isochrone of pre-main-sequence evolutionary tracks to 
obtain $T_{\rm eff}$ and $M_{\star}$. For $L_{\star}\,{<}\,0.4\,L_{\odot}$ (approximately for $T_{\rm eff}\,{<}\,3900\,{\rm K}$), we adopt the models presented by \citet{Baraffe2015}, whereas we take the nonmagnetic evolutionary models by \citet{Feiden2016} for $L_{\star}\,{\geq}\,0.4\,L_{\odot}$. This procedure is motivated 
by the fact that the two sets of isochrones overlap well around $T_{\rm eff}\,{\sim}\,3900\,{\rm K}$. Once the stellar parameters are given, we use the corresponding \texttt{BT-Settl} models to represent the atmospheric spectra \citep{Allard2011}. In the grid, there are three points for $\Psi$: 1.05, 1.15, 1.25. We consider three values for $h_{100}$: 5\,AU, 10\,AU and 15\,AU. For $R_{c}$, we also sample three values: 10\,AU, 30\,AU and 100\,AU. We consider three disk-to-stellar mass ratios, i.e., 0.001, 0.01 and 0.1. Then, $M_{\rm dust}$ for each model is determined by assuming a gas-to-dust mass ratio of 100.  

\begin{figure}[!t]
\centering
  \includegraphics[width=0.40\textwidth]{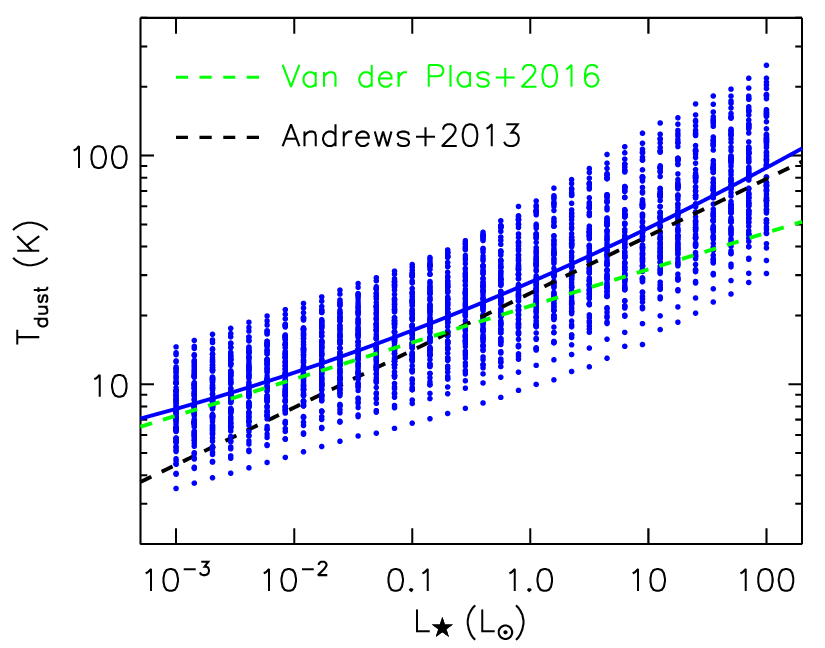}
  \caption{Mass-averaged dust temperature as a function of stellar luminosity. The blue solid curve, expressed as 
     ${\rm log }\,T_{\rm dust}\,{=}\,1.445\,{+}\,0.224\,{\rm log}\,L_{\star}\,{+}\,0.013\,({\rm log}\,L_{\star})^2$  
	 is the best-fit relation to our models. The black dashed line refers to the relation $T_{\rm dust}\,{=}\,25\,(L_{\star}/L_{\odot})^{0.25}\,{\rm K}$ presented by \citet{Andrews2013}. The green dashed line shows the relation $T_{\rm dust}\,{=}\,22\,(L_{\star}/L_{\odot})^{0.16}\,{\rm K}$ suggested by \citet{vanderPlas2016}.}
\label{fig:lumtdust}
\end{figure}

Figure~\ref{fig:lumtdust} shows the mass-averaged dust temperature $T_{\rm dust}$ as a function of $L_{\star}$. Although there is a large dispersion in $T_{\rm dust}$ 
for each $L_{\star}$ bin, an obvious trend is observed, with cooler disks around fainter stars. Moreover, the decreasing tendency of $T_{\rm dust}$ with 
$L_{\star}$ appears steeper for stars with $L_{\star}\,{\gtrsim}\,0.2\,L_{\odot}$ than that for systems with lower luminosities. The same phenomenon is also seen 
in Figure 17 of \citet{Andrews2013}, in which the dust temperature decrease seems to flatten out toward the low luminosity regime. 
We fit a polynomial (second order) to the correlation between $T_{\rm dust}$ and $L_{\star}$. The best fit is represented by 

\begin{equation}
{\rm log }\,T_{\rm dust}\,{=}\,1.445\,{+}\,0.224\,{\rm log}\,L_{\star}\,{+}\,0.013\left({\rm log}\,L_{\star}\right)^2. 
\label{eqn:tdust}
\end{equation}

Our relation for $L_{\star}\,{<}\,0.2\,L_{\odot}$ predicts similar dust temperature to the result given by \citet{vanderPlas2016}, whereas for $L_{\star}\,{\gtrsim}\,0.2\,L_{\odot}$ we obtain similar dust temperature to the prescription of \citet{Andrews2013}. The small difference between our and literature results is understood because of the difference in the choice of dust density distribution, disk parameters and dust opacities. For instance, we used porous dust opacities in the modeling, while other studies adopted compact dust opacities. Moreover, whether or not taking into account the stellar-mass dependent disk outer radius \citep{Andrews2018b, Andrews2020} and interstellar radiation will also influence the heating of disks as a function of spectral type, and therefore affect the $T_{\rm dust}\,{-}\,L_{\star}$ relation \citep[e.g.,][]{vanderPlas2016,Hendler2017}.  

\begin{figure}[!t]
\centering
  \includegraphics[width=0.40\textwidth]{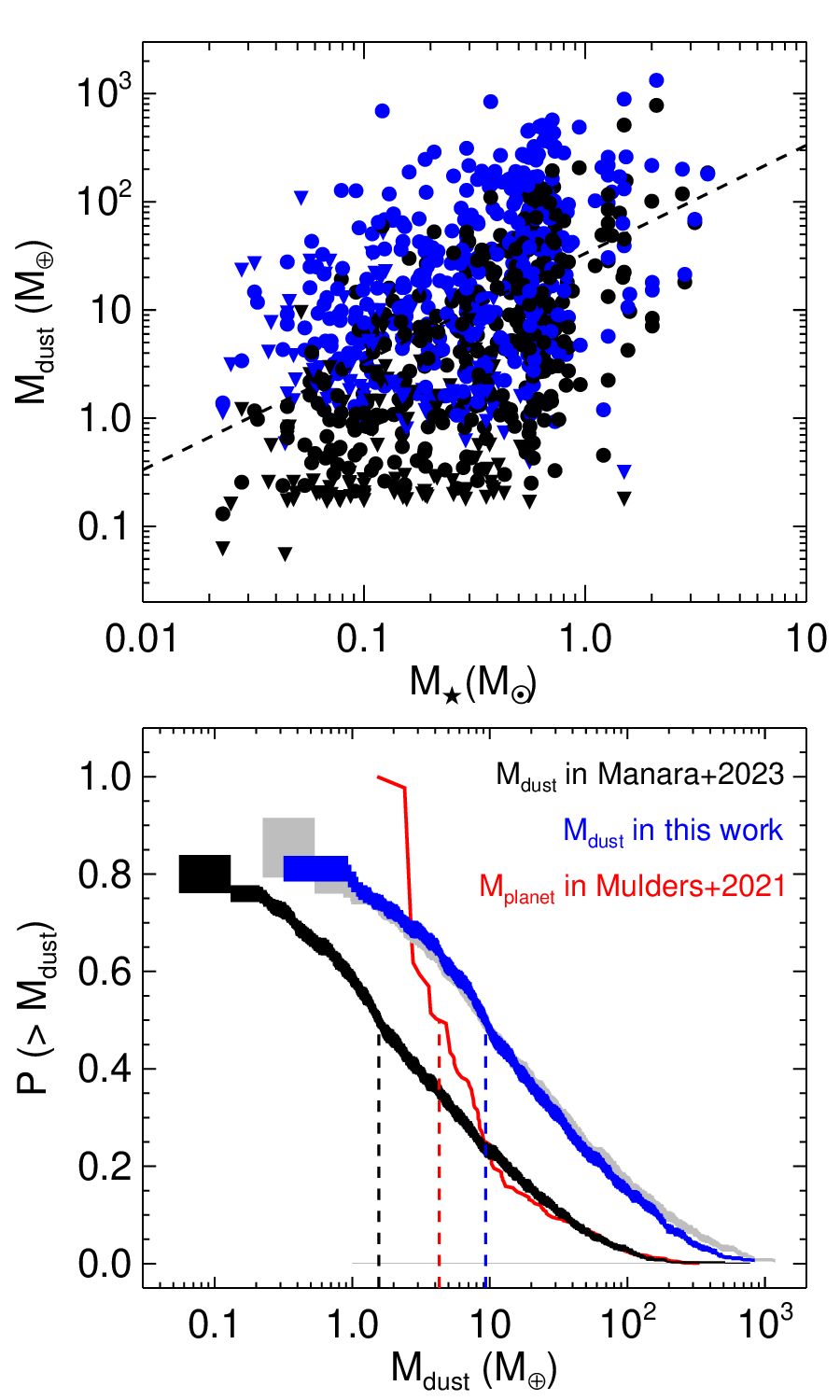}
  \caption{Dust masses in protoplanetary disks and their statistics. {\it Upper panel:} $M_{\rm dust}$ as a function of $M_{\star}$. Black symbols show the results derived by \citet{Manara2023}, while blue symbols represent our calculations considering porous dust opacities and a $T_{\rm dust}-L_{\star}$ scaling given by Eq.~\ref{eqn:tdust}, see Sect.~\ref{sec:exoplanet}. Millimeter detections are indicated with dots, whereas triangles mean upper limits of the millimeter flux are reported. The dashed line depicts the value of $100{\cdot}M_{\rm dust}\,{=}\,0.01{\cdot}M_{\star}$. {\it Bottom panel:} CDF of $M_{\rm dust}$. The grey curve shows the result calculated with porous dust opacities and a constant dust temperature $T_{\rm dust}\,{=}\,20\,\rm{K}$. The red curve refers to the distribution of exoplanet masses obtained by \citet{Mulders2021}. The vertical dashed lines mark the median values of each distribution.}
\label{fig:mstarmdust}
\end{figure}

To calculate $M_{\rm dust}$, we directly took $L_{\star}$, $F_{\nu}$ and $D$ of each target provided by \citet{Manara2023}. When $L_{\star}$ is 
not available, we set $T_{\rm dust}\,{=}\,20\,\rm{K}$. The measured flux density at $1.3\,\rm{mm}$ is used with a higher priority since the optical depth is lower at longer wavelengths. The mass absorption coefficients of porous grains for the LGP are adopted in the calculation, i.e., $\kappa_{\rm 1.3mm}\,{=}\,0.37\,\rm{cm^2/g}$ and $\kappa_{\rm 0.88mm}\,{=}\,0.84\,\rm{cm^2/g}$. As can be seen in the upper panel of Figure~\ref{fig:mstarmdust}, adopting porous dust opacities results in a systematically higher $M_{\rm dust}$ than that calculated by assuming $\kappa_{\nu}\,{=}\,2.3(\nu/230\,\rm{GHz)\,cm^2/g}$ and a constant $T_{\rm dust}\,{=}\,20\,\rm{K}$. 
Moreover, we used the Kaplan-Meier estimator from the \texttt{ASURV} package to estimate the CDFs and median dust masses while
properly accounting for the upper limits \citep{Feigelson1985,Isobe1986}. The median values are $1.6\,M_{\oplus}$ and $9.3\,M_{\oplus}$ for 
$M_{\rm dust}$ calculated by \citet{Manara2023} and in this study, respectively, which means that dust masses in protoplanetary disks may be underestimated by a factor of ${\sim}\,6$ if the grains are in fact porous and not compact. For comparison, the grey curve in the bottom panel of Figure~\ref{fig:mstarmdust} shows the CDF of $M_{\rm dust}$ calculated with the porous dust opacities and a constant dust temperature $T_{\rm dust}\,{=}\,20\,\rm{K}$. The distribution is similar to that derived by scaling $T_{\rm dust}$ with $L_{\star}$. This is understood because the disks have a wide range of stellar luminosity $10^{-3}\,L_{\odot}\,{\lesssim}\,L_{\star}\,{\lesssim}\,100\,L_{\odot}$, and the effect of dust temperature on the distribution smooths out when the analysis is performed in a large sample scale. \citet{Mulders2021} analyzed the mass distribution of the exoplanet population detected from the Kepler transit survey \citep[e.g.,][]{Borucki2010,Thompson2018} and radial velocity surveys from \citet{Mayor2011}, because both surveys have a well-characterized detection bias. The red curve in the bottom panel of Figure~\ref{fig:mstarmdust} refers to the CDF of the exoplanet mass when accounting for observational selection and detection biases. The median mass of the exoplanets is $4.2\,M_{\oplus}$ that is about two times lower than the median dust mass calculated in this work. Moreover, we do not see any exoplanetary systems with masses higher than the most massive disks. Our study shows that, if dust grains in disks are porous, the problem of insufficient mass for planet formation raised in the ALMA era may be resolved.

In the above analysis, we adopt an effective maximum grain size of 1\,mm. The grain size has a direct impact on the dust opacity, and is commonly investigated via the millimeter spectral index \citep[e.g.,][]{Andrews2005,Ricci2010b}. Future observations at multiple (sub-)millimeter wavelengths are needed to constrain the grain size that is essential for a better characterization of the dust mass distribution. To observationally prove that dust grains are porous and constrain the porosity $\mathcal{P}$, multiwavelength observations and analysis are mandatory. 

Grain porosity alters the absorption, scattering and polarization behavior of dust grains \citep[e.g.,][]{Semenov2003,Ysard2018}, and therefore influences the observable appearance of protoplanetary disks. The position, strengths, width, and shape of dust emission features in the mid-infrared domain depend on the grain structure \citep[e.g.,][]{Voshchinnikov2008,Vaidya2011}. Detailed decomposition of spectra obtained with the Infrared Spectrograph (IRS) on board Spitzer has demonstrated that the emission bands of forsterite and enstatite are best matched with porous grains, rather than compact spherical grains \citep[e.g.,][]{Bouwman2008,Juhasz2010}. Analyzing the spectra from the Mid-Infrared Instrument (MIRI) equipped on JWST gives the same conclusion \citep[e.g.,][]{Jang2024}. The accumulating MIRI spectra will shed more insights into the dust properties particularly in faint disks that were not accessible by Spitzer.   

\citet{Kirchschlager2014} found that grain porosity strongly influences the scattered-light maps of protoplanetary disks. For highly tilted disks (e.g., $i\,{\gtrsim}\,75^{\circ}$), the flux of the scattered light at optical wavelengths is significantly enlarged when porous dust opacities are included. Once the disk deviates from a face-on orientation (e.g., $i\,{\gtrsim}\,5^{\circ}$), the scattered light maps show a dark lane that is characteristic for inclined disks. The dark lane appears more pronounced assuming compact dust opacities. By analyzing scattered-light maps from radiative transfer simulations, \citet{Tazaki2019a} found that porous dust aggregates large compared to near-infrared wavelengths show marginally grey or slightly blue in total or polarized 
intensity, while large compact dust grains give rise to reddish scattered-ligh colours in total intensity. High resolution near-infrared observations can be used to verify the model predictions \citep[e.g.,][]{Fukagawa2010,Avenhaus2018}.

\begin{figure*}[!t]
\centering
  \includegraphics[width=0.8\textwidth]{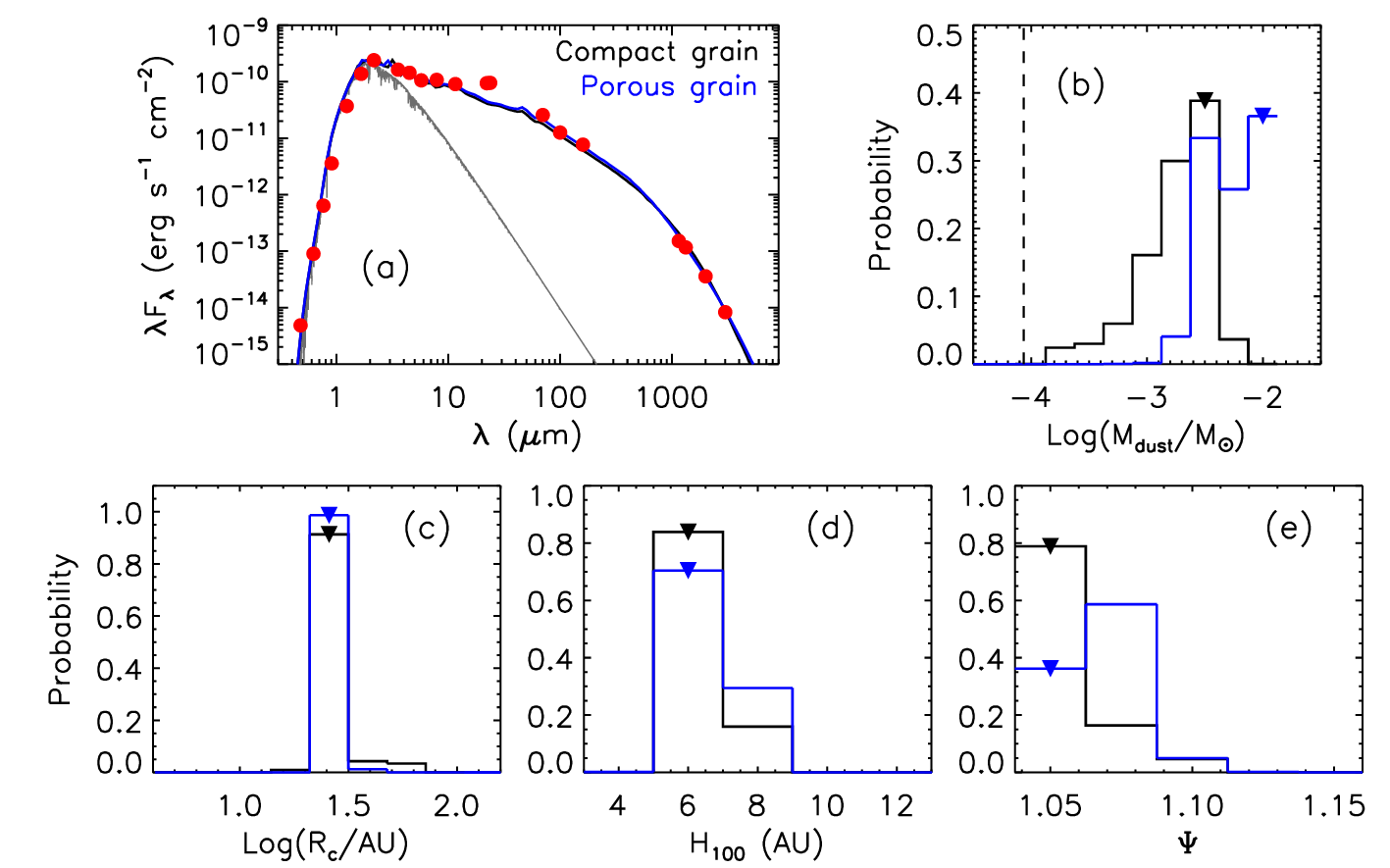}
  \caption{Fitting results of the IRAS\,04370+2559 disk. {\it Panel (a):} SEDs of IRAS\,04370+2559. The best-fit radiative transfer models with compact dust 
          opacities and porous dust opacities are shown with black lines and blue lines, respectively. The grey curve refers to the input \texttt{BT-Settl} spectrum, and 
		  photometric data points are overlaid with red dots. {\it Panels (b)$-$(e):} Bayesian probability distributions for ${\rm Log}_{10}(M_{\rm dust}/M_{\odot})$, 
          ${\rm Log}_{10}(R_{\rm c}/{\rm AU})$, $H_{100}$ and $\Psi$. The triangles indicate the best-fit parameter values. The vertical dashed line in panel (b) marks 
		  the analytic dust mass calculated with Eq.~\ref{eqn:mana} by assuming $T_{\rm dust}\,{=}\,20\,\rm{K}$ and $\kappa_{1.3\rm{mm}}\,{=}\,2.3\,\rm{cm^2/g}$.}
\label{fig:sedprob}
\end{figure*}

There exist significant differences between polarimetric images of disks composed of porous particles and compact spheres \citep{Min2012}. Simulated polarization maps at optical wavelengths 
reveal an increase in the polarization degree by a factor of about four when porous grains are considered \citep{Kirchschlager2014}. The wavelength dependent polarization reversal (i.e., $90^{\circ}$ flip of the polarization direction at visible light) depends strongly on the grain porosity, and thus has diagnostic potential for dust properties. \citet{Tazaki2019} investigated how the grain structure and porosity alter polarimetric images at millimeter wavelengths. The polarization pattern of disks containing moderately porous particles show near- and far-side asymmetries at (sub-)millimeter wavelengths. Porous grains exhibit a weaker wavelength dependence of scattering polarization than solid spherical grains. \citet{Brunngraber2021} investigated the effect of grain porosity on the polarization degree due to self-scattering in the (sub-)millimeter wavelength range. They found that porous dust grains with moderate filling factors of about 10\% increase the degree of polarization by up to a factor of four compared to compact grains. However, the degree of polarization drops rapidly when the porosity is very high, with a filling factor of 1\% or lower,  because of the low opacity and optical depth. These theoretical studies show that multi-wavelength polarimetric observations are crucial to constrain the structure and porosity of dust grains in protoplanetary disks \citep[e.g.,][]{Zhang2023,Ueda2024}. 

\section{Modeling the SED of IRAS 04370+2559}
\label{sec:iras04370}

Protoplanetary disks, especially in the inner region and dense midplane, are likely to be optically thick at millimeter 
wavelengths \citep[e.g.,][]{Wolf2008,Pinte2016,Liuy2019,Ueda2020}, which means that $M_{\rm dust}$ estimated using Eq.~\ref{eqn:mana} is underestimated. 
Therefore, in reality the CDF of $M_{\rm dust}$ should shift towards the higher abscissa value in Figure~\ref{fig:mstarmdust}. Self-consistent radiative 
transfer modeling is an appropriate way to treat the optical depth effect, and therefore better constrain the dust mass. In this section, we conduct a 
detailed modeling for the SED of the IRAS\,04370+2559 disk, and showcase the difference in $M_{\rm dust}$ derived using different approaches and dust opacities.    

\subsection{Build the observed SED with new IRAM/NIKA-2 observations}
\label{sec:obssed}

IRAS\,04370+2559 is a T Tauri star located in the Taurus star formation region at a distance of $D\,{=}\,137.4\,\rm{pc}$ \citep{Gaia2023}. Besides the stellar 
properties of IRAS\,04370+2559 is representative for T Tauri stars \citep{Gullbring1998}, there are two reasons why we select it for the test. First, a large number 
of photometric data points from optical to millimeter regimes are available either in the archive or from our new IRAM/NIKA-2 observations. Second, the observed 
SED of IRAS\,04370+2559, shown with red dots in panel (a) of Figure~\ref{fig:sedprob}, displays a flux drop at $\lambda\,{\sim}\,70\,\mu{\rm m}$. This is a strong indication of dust settling \citep{Liuy2012,Dullemond2004,DAlessio2006}, which implies that a huge amount of material is very likely concentrated close to the disk midplane, creating high optical depths. In this case, a detailed modeling of the SED is necessary to constrain the dust mass, and it is expected to have a large difference in $M_{\rm dust}$ derived with different approaches. To compile the SED, we collect photometry from the Sloan Digital Sky Survey \citep{Adelman-McCarthy2011}, 2MASS catalog \citep{Skrutskie2006}, 
Spitzer/IRAC and MIPS measurements \citep{Luhman2010,Rebull2010} and Herschel/PACS data \citep{Ribas2017}. In the millimeter, there are 
data points at 1.3\,mm and 3\,mm obtained with the Submillimeter Array \citep{Andrews2013} and Combined Array for Research in 
Millimeter-wave Astronomy \citep{Babaian2020}, respectively.

We conduct new IRAM/NIKA-2 observations towards the IRAS 04370+2559 disk. Details about the observation, data reduction and flux extraction 
are described in Appendix~\ref{sec:iram}. The measured flux densities at 1.1\,mm and 2\,mm are $58.2\,{\pm}\,2.4\,\rm{mJy}$ and $23.9\,{\pm}\,0.6\,\rm{mJy}$, respectively. 
These efforts allow us to build the SED with an excellent wavelength coverage, see panel (a) of Figure~\ref{fig:sedprob}. 
\citet{Andrews2013} fitted the optical part of the SED, and derived $T_{\rm eff}\,{=}\,3778\,\rm{K}$ and $L_{\star}\,{=}\,0.9\,L_{\odot}$ assuming a distance 
of 140\,pc. We rescale their stellar luminosity by adopting the Gaia distance of $D\,{=}\,137.4\,\rm{pc}$ \citep{Gaia2023}, yielding $L_{\star}\,{=}\,0.86\,L_{\odot}$. 
Moreover, an extinction of $A_{V}=10.65\,\rm{mag}$ is required to reproduce the optical photometry. We use the extinction law of \citet{Cardelli1989} with $R_{V}\,{=}\,3.1$. 

\subsection{Results}
\label{sec:sedana}

Using the model grid generated in Sect.~\ref{sec:rtmodel}, we evaluate the quality of fit to the SED by using the $\chi^2_{\rm SED}$ metric. 
We also compare the expected effective disk radius $R68_{\rm exp}$ of IRAS\,04370+2559 with the $R68$ value of each model, because the disk size 
has an important influence on the derived dust mass \citep{Ballering2019,Liuy2022}. First, an expected 0.88\,mm flux is derived by 
extrapolating the millimeter part of the observed SED. Then, we calculate $R68_{\rm exp}$ using the \citet{Hendler2020} relation  
\begin{equation}
{\rm log}R68\,{=}\,(2.16 \pm 0.11)\,{+}\,(0.53 \pm 0.12)\,{\rm log}F_{0.88{\rm mm}}.
\label{eqn:r68fnu}
\end{equation}
The derived $R68_{\rm exp}\,{=}\,46.6\,{\pm}\,16.7\,\rm{AU}$ is shown with the red dot in the bottom panel of Figure~\ref{fig:r68fnu}. 
A comparison of $R68$ between the observation and model predictions gives an extra term to the total fit metric 
\begin{equation}
 \chi^2 = \chi^2_{\rm SED} + g\,\chi^2_{R68}.	
\end{equation}  
The $g$ factor is used to balance the weighting between both observables, and it is determined by comparing the median $\chi^2_{\rm SED}$ with median $\chi^2_{R68}$
of all the models. 

We conduct a Bayesian analysis by assuming flat priors for each parameter. The relative probability of a model in the parameter space is 
given by ${\rm exp}(-\chi^2/2)$ \citep[e.g.,][]{Lay1997,Pinte2008}. Then, the marginalized probability distribution for each parameter is obtained 
by first adding the individual probabilities of all the models with a common value of the parameter, and then normalizing to the total probability for that 
parameter. The resulting marginalized probability distributions are presented in panels (b)$-$(e) of Figure~\ref{fig:sedprob}. The triangles indicate the 
parameter values of the best-fit model, most of which are identical to the most probable values.

The best-fit model with porous dust opacities have a dust mass of $0.01\,M_{\odot}$ that is about 3 times larger than the best-fit 
value $3.2\,{\times}\,10^{-3}\,M_{\odot}$ when including compact dust opacities. To calculate the analytic dust mass using Eq.~\ref{eqn:mana}, we adopt
$\kappa_{\rm 1.3mm}\,{=}\,2.3\,\rm{cm^2/g}$ and $T_{\rm dust}\,{=}\,20\,\rm{K}$ for consistency with literature works. 
We take $F_{1.3\,\rm{mm}}\,{=}\,51.7\,\rm{mJy}$ \citep{Andrews2013}, and derive $M_{\rm dust.ana}\,{=}\,8.6\,{\times}\,10^{-5}\,M_{\odot}$, which is 
indicated with the vertical dashed line in panel (b) of Figure~\ref{fig:sedprob}. The dust masses from radiative transfer modeling with porous grains 
and compact grains are 116 and 37 times higher than the analytic dust mass, respectively. In order to disentangle the effects of dust porosity and 
running radiative transfer modeling on the difference in the mass determination, we further calculate the analytic dust mass 
using $\kappa_{\rm 1.3mm}\,{=}\,0.37\,\rm{cm^2/g}$ of porous grains for the LGP. The derived mass $5.4\,{\times}\,10^{-4}\,M_{\odot}$ is 
about 6 times higher than that calculated with $\kappa_{\rm 1.3mm}\,{=}\,2.3\,\rm{cm^2/g}$. Comparing this factor with 116 indicates that running radiative transfer
models is mostly responsible for the large difference. We note that whether dust porosity or radiative tranfer modeling contributes more to the difference in the mass determination 
varies from source to source, as previous studies show that the factor of mass underestimation induced by radiative transfer modeling ranges from a few to hundreds 
depending on the optical depth of the disk \citep{Ballering2019,Liuy2022}.

The significant difference in $M_{\rm dust}$ between the traditionally analytic calculation and radiative transfer modeling is due to two main reasons.
First, the best-fit model features a low flaring index ($\Psi\,{=}\,1.05$) and scale height ($h_{100}\,{=}\,6\,\rm{AU}$), leading to a low 
mass-averaged dust temperature $T_{\rm dust}\,{\sim}\,14\,\rm{K}$. Second, a small characteristic radius $R_{\rm c}\,{=}\,25.8\,\rm{AU}$ is required 
to reproduce the expected effective radius $R68_{\rm exp}$. The narrow radial density distribution creates highly optically thick regions, 
resulting in a large underestimation of $M_{\rm dust}$. We further use the 3\,mm flux $F_{3\,\rm{mm}}\,{=}\,8.3\,\rm{mJy}$ \citep{Babaian2020} 
and $\kappa_{\nu}\,{=}\,2.3(\nu/230\,\rm{GHz)\,cm^2/g}$ to calculate the analytic dust mass. The result is $M_{\rm dust.ana}\,{=}\,1.4\,{\times}\,10^{-4}\,M_{\odot}$, roughly 
2 times larger than that derived with the 1.3\,mm data. This implies that the IRAS\,04370+2559 disk is likely optically thick at 1.3\,mm. Future spatially resolved 
observations at longer wavelengths are required to constrain the disk size and dust density distribution that allow a better measurement of 
the dust mass \citep[e.g.,][]{Macias2021,Guidi2022,Guerra-Alvarado2024}. 

\section{Summary}
\label{sec:summary}

The total dust mass in protoplanetary disks is an important parameter that evaluates the potential for planet formation. 
The determination of dust mass is dependent on the choice of dust properties. There is growing evidence that dust grains in protoplanetary 
disks might be porous. In this work, using self-consistent radiative transfer models, we compared the dust temperature $T_{\rm dust}$ and 
emergent millimeter fluxes between models incorporating compact dust opacities and porous dust opacities. 

The results show that $T_{\rm dust}$ is similar between models with the two types of dust opacities, but porous dust grains 
yield systematically lower millimeter fluxes than compact dust grains because of the lower emissivity. This fact has a 
pronounced impact on the solid mass budget for planet formation. We re-calibrated the $T_{\rm dust}\,{-}\,L_{\star}$ relation 
for a wide range of stellar mass, and obtained a second-order polynomial in the explored stellar mass regime. We calculated the 
analytic dust masses of a large sample of disks with our $T_{\rm dust}\,{-}\,L_{\star}$ relation and porous dust opacities. 
The median dust mass $9.3\,M_{\oplus}$ is about 6 times higher than the literature result $1.6\,M_{\oplus}$, and it is  
about 2 times higher than the median mass of exoplanets investigated by \citet{Mulders2021}. A comparison of the CDF between disk dust 
masses and exoplanet masses shows that there are no exoplanetary systems with masses higher than the most massive disks, if dust grains 
in disks are in fact porous. Our study suggests that adopting porous dust opacities may alleviate the problem of insufficient dust solids for planet formation. 

Mass calculation using the traditionally analytic approach always underestimates the true value due to the optical depth effect. 
We took the IRAS\,04370+2559 disk as an example to show a combined effect of optical depth and porous dust opacities on the mass 
estimation. We conducted new IRAM/NIKA-2 observations towards IRAS\,04370+2559, enabling a better wavelength sampling of the 
observed SED. A large grid of radiative transfer models are fitted to the observation. The best-fit dust mass is about 100 times 
larger than the analytic dust mass when the porous dust opacities are adopted in the modeling. 

\begin{acknowledgements}
We thank the anonymous referee for the constructive comments that highly improved the manuscript. YL acknowledges financial supports by the Natural Science Foundation of China (Grant No. 11973090), and by the International Partnership Program of Chinese Academy of Sciences (Grant number 019GJHZ2023016FN). This work is based on observations carried out under project number 040-17 with the IRAM 30m telescope. IRAM is supported by INSU/CNRS (France), MPG (Germany) and IGN (Spain). We would like to thank the IRAM staff for their support during the NIKA2 campaigns. The NIKA2 dilution cryostat has been designed and built at the Institut Néel. We acknowledge the crucial contribution of the Cryogenics Group, and in particular Gregory Garde, Henri Rodenas, Jean Paul Leggeri, Philippe Camus. This work has been partially funded by the Foundation Nanoscience Grenoble and the LabEx FOCUS ANR-11-LABX-0013 and is supported by the French National Research Agency under the contracts “MKIDS”, “NIKA” and ANR-15-CE31-0017 and in the framework of the “Investissements d’avenir” program (ANR-15-IDEX-02). This work has benefited from the support of the European Research Council Advanced Grant ORISTARS under the European Union’s Seventh Framework Programme (Grant Agreement no. 291294). The NIKA2 data were pre-processed and calibrated with the NIKA2 collaboration pipeline. This research has made use of Herschel data; Herschel is an ESA space observatory with science instruments provided by European-led Principal Investigator consortia and with important participation from NASA. FK has received funding from the European Research Council (ERC) under the European Union’s Horizon 2020 research and innovation programme DustOrigin (ERC-2019-StG-851622). In this study, a cluster is used with the SIMT accelerator made in China. The cluster includes many nodes each containing 4 CPUs and 8 accelerators. The accelerator adopts a GPU-like architecture consisting of a 16GB HBM2 device memory and many compute units. Accelerators connected to CPUs with PCI-E, the peak bandwidth of the data transcription between main memory and device memory is 16GB/s. This work is partially supported by cosmology simulation database (CSD) in the National Basic Science Data Center (NBSDC-DB-10). 
\end{acknowledgements}

\bibliographystyle{aa}
\bibliography{dustmass.bib}

\begin{appendix}

\section{IRAM/NIKA-2 observations of the IRAS\,04370+2559 disk}
\label{sec:iram}	

In this section, we present the details about the observations and data reduction of the IRAM/NIKA-2 observations towards IRAS\,04370+2559. The on-the-fly observations were conducted simultaneously at 1.25 and 2\,mm as part of the IRAM project 040-17 on October 27, 2017 in four consecutive scans, for a total duration of 8.9\,min. The mean elevation of the source was $73^{\circ}$ and the mean opacity at this elevation was on the order of 0.13 at 2\,mm and 0.32 at 1.25\,mm\,. The sky rms was low and stable.

The data were processed with the \textit{Scanam\_nika} package, a branch of \textit{Scanamorphos} adapted to NIKA-2 data \citep{Roussel2013}. The underlying principle is the same as for Herschel data, i.e. the full exploitation of the redundancy built in the observations to subtract the atmospheric and instrumental noise from the data. More details are given in \citet{Ejlali2024}. The flux calibration established by \citet{Perotto2020} is refined based on aperture photometry on more than a hundred finely-sampled maps of Uranus observed from October 2017 to January 2023. This leads to beam solid angles of $(188\,{\pm}\,11)$ and $(381\,{\pm}\,11)$ square arcseconds at 1.15 and 2\,mm, respectively.

\begin{figure} 
   \centering
   \includegraphics[width=0.22\textwidth]{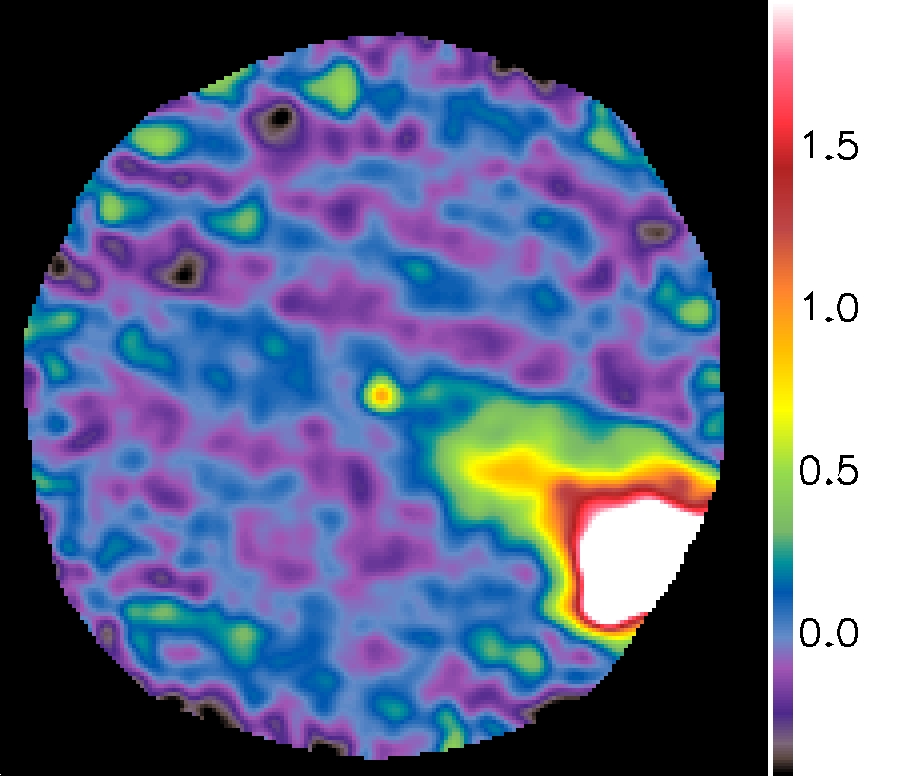}
   \includegraphics[width=0.22\textwidth]{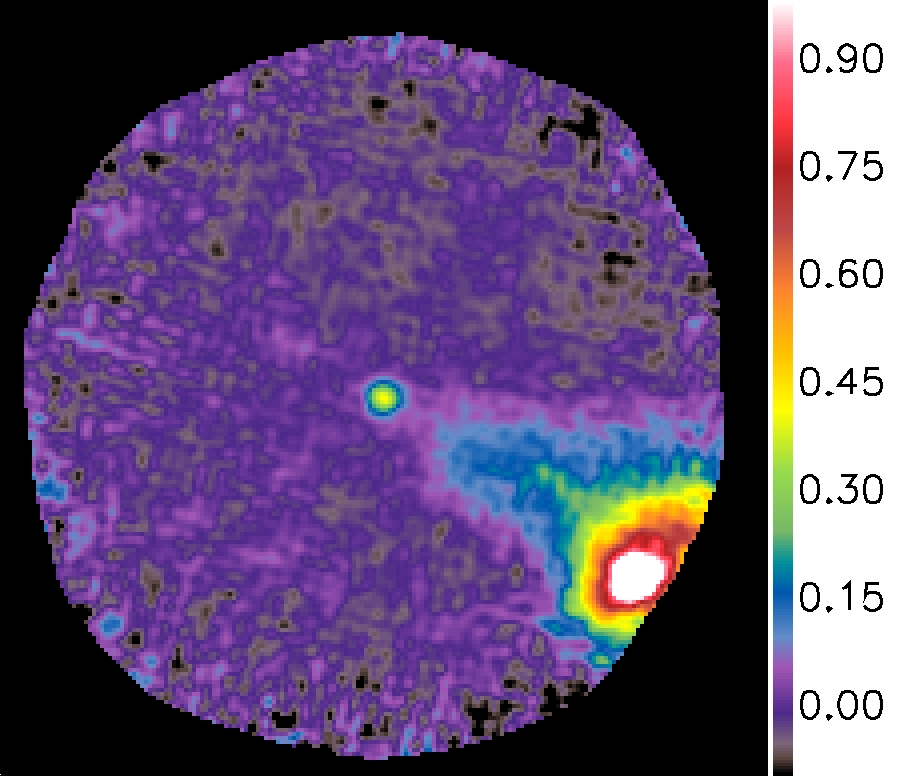}
   \caption{NIKA2 maps of IRAS\,04370+2559 at 1.1\,mm (left) and 2.0\,mm (right) in mJy/pixel. The pixel size is $3^{\prime \prime}$ and the data have been convolved to a FWHM resolution of 20$^{\prime \prime}$.} 
   \label{fig:nika2_images}
\end{figure}

\begin{figure} 
   \centering
   \includegraphics[width=0.15\textwidth]{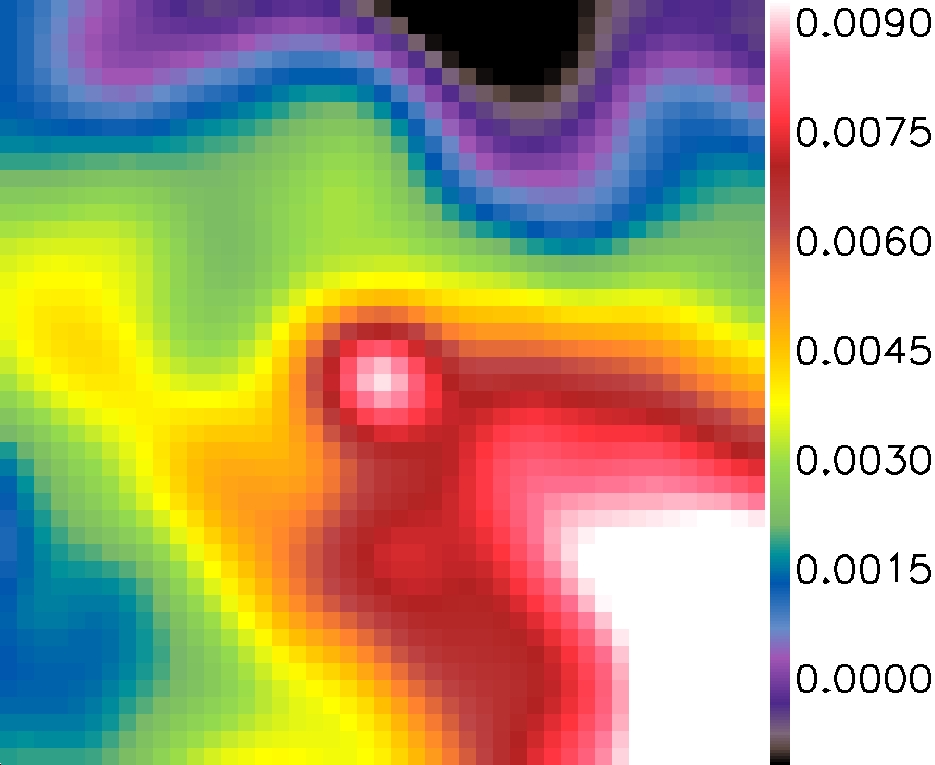}
   \includegraphics[width=0.15\textwidth]{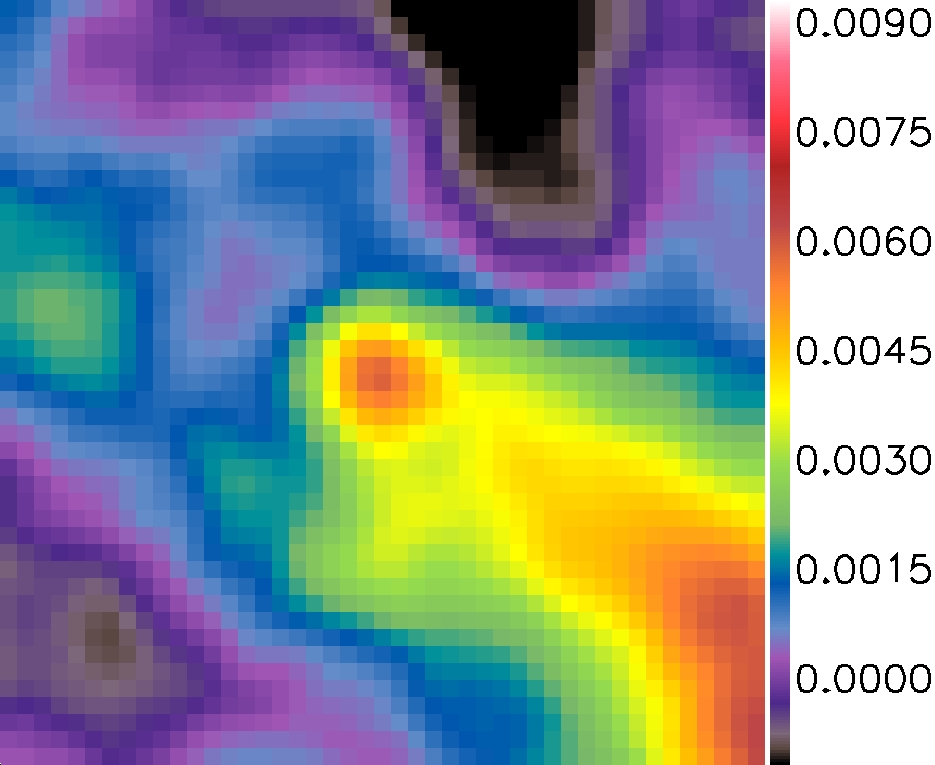}
   \includegraphics[width=0.15\textwidth]{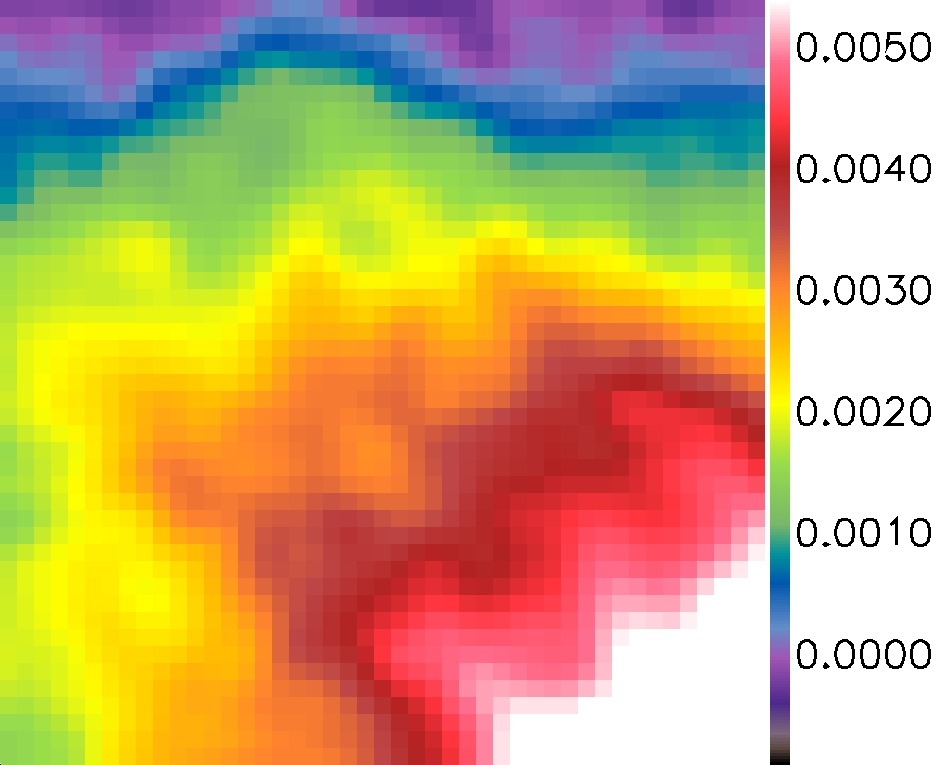}   
   \caption{Simulation of the effect of the 2\,mm noise on the data: zoom on the input 160-$\mu$m map (left), on the simulation output (middle) and on the difference between the two maps, with a smaller dynamic range (right). The pixel size and FWHM resolution are the same as in Figure~\ref{fig:nika2_images}, and the flux unit is Jy/pixel.}
   \label{fig:nika2_sim160}
\end{figure}

The maps are shown in Figure~\ref{fig:nika2_images}. To assess the reliability of the photometry, simulations were performed, using as input the longest-wavelength infrared data shortward of 1.15\,mm where the source is detected, i.e. the PACS 160\,$\mu$m map. Time series were simulated from this map as if they had been obtained with NIKA-2, with the same observation geometry and sampling. They were rescaled so that the median signal-to-noise ratio for the source was identical in the simulation and in the NIKA-2 observations. The noise extracted from NIKA-2 data was added, and then the resulting time series were processed in the same fashion as the real data, and the final map was rescaled back to the original unit. The results for the 2\,mm noise are shown in Figure~\ref{fig:nika2_sim160}. They confirm that while the recovery of extended emission is challenging in such small and shallow maps, the source of interest is preserved, since no residual is visible at its location in the difference map between the simulation input and output.

Photometric information from the NIKA2 maps was gathered via aperture photometry. We used an aperture radius of 15$''$ and subtracted a local background estimated in an annulus between $30\,{-}\,50\prime\prime$, avoiding the extended emission associated with the bright source to the south-west (i.e., L\,1527\,IRS). We corrected these fluxes by aperture correction factors estimated from the above-mentioned Uranus observations. We arrive at the following flux densities:\\
 $F_\nu$(1.1\,mm) = 58.2 mJy $\pm$ 2.4 mJy $\pm$ 3.5 mJy, \\
 $F_\nu$(2.0\,mm) = 23.9 mJy $\pm$ 0.6 mJy $\pm$ 0.8 mJy. \\
The first error term is derived from the random noise in the photometry aperture via $\sigma_{\rm{(flux \, density)}}\,{=}\,\sqrt{N_{\rm pix}}\,{\times}\,\sigma_{\rm map}$, with $N_{\rm pix}$ denoting the number of pixels in the aperture. The second term assumes formal uncertainties of 3.5\% and 6.0\% in the general flux calibration for 2.0\,mm and 1.25\,mm, respectively. These numbers are derived from the variations in the Uranus fluxes during the observing run containing our target observations, and are in line with the typical values reported in \citet{Perotto2020}.  

\end{appendix}

\end{document}